**Perpetual Observational Studies: New strategies to support efficient implementation of observational studies and randomized trials in infectious diseases**

N. Hassoun-Kheir MD, Infection Control Program, Geneva University Hospitals and Faculty of Medicine, Geneva Switzerland

C.H. van Werkhoven PhD MD, Julius Center for Health Sciences and Primary Care, University Medical Center Utrecht, Utrecht University, the Netherlands

J. Dunning PhD MD, Pandemic Sciences Institute, University of Oxford, Oxford, UK

T. Jaenisch PhD MD, Heidelberg Institute of Global Health (HIGH), Heidelberg University Hospital, Heidelberg, Germany

J. van Beek PhD, Department of Viroscience, Erasmus MC, University Medical Center Rotterdam, Rotterdam, The Netherlands

J. Bielicki PhD MD, Institute for Infection and Immunity, St George's, University of London, London, UK; Infection Prevention and Control, Universitäts-Kinderspital beider Basel, Basel, Switzerland

C. Butler Prof, Nuffield Department of Primary Care Health Sciences, Oxford University

B. Francois MD, Medical-Surgical Intensive Care Unit and Inserm CIC 1435 & UMR 1092, CHU Limoges, Limoges, France

S. Harbarth Prof, Infection Control Program, Geneva University Hospitals and Faculty of Medicine, Geneva Switzerland

A. C. Hernandez Padilla MD, Inserm CIC 1435 and Medical-Surgical Intensive Care Unit, CHU Dupuytren, Limoges, France

P. Horby Prof, Pandemic Sciences Centre, University of Oxford, Oxford, UK

M. Koopmans Prof, Department of Viroscience, Erasmus MC, University Medical Center Rotterdam, Rotterdam, The Netherlands, and Pandemic and Disaster Preparedness Research Centre, Rotterdam, The Netherlands

J. Lee BSc, Pandemic Sciences Institute, University of Oxford, Oxford, UK

J. Rodriguez-Baño Prof, Infectious Diseases and Microbiology, Hospital Universitario Virgen Macarena, Department of Medicine, University of Sevilla, Biomedicine Institute of Sevilla (IBiS), CSIC, Sevilla, Spain and, Centro de Investigación Biomédica en Red en Enfermedades Infecciosas (CIBERINFEC), Instituto de Salud Carlos III, Madrid, Spain

E. Tacconelli Prof, Dept. of Infectious Diseases, University of Verona, Verona, Italy

Y. Themistocleous MBChB, Pandemic Sciences Institute, University of Oxford, Oxford, UK

A.W. van der Velden PhD, Julius Center for Health Sciences and Primary Care, University Medical Center Utrecht, Utrecht, the Netherlands

M. Bonten Prof, Julius Center for Health Sciences and Primary Care, University Medical Center Utrecht, Utrecht University, the Netherlands

H. Goossens Prof, Laboratory of Medical Microbiology, Vaccine & Infectious Disease Institute, University of Antwerp, Antwerp, Belgium

M.E.A. de Kraker PhD, Infection Control Program, Geneva University Hospitals and Faculty of Medicine, Geneva Switzerland

on behalf of the ECRAID-Base consortium

**Corresponding author:**

Marlieke E.A. de Kraker PhD

Infection Control Program, Geneva University Hospitals and Faculty of Medicine

Gabrielle-Perret-Gentil 4, 1205 Geneva, Switzerland

Phone: +41-223723369, email: marlieke.dekraker@hcuge.ch

**Background**

Emerging infectious diseases are a growing threat, through population growth, increased trade and travel, urbanization, deforestation, and climate change [1]. Clinical research in response to emerging infectious diseases is challenging; selecting and contracting appropriate study sites for a trial is time-consuming, and often too few patients can be timely recruited to acquire high-quality evidence about the best treatment strategies [2]. In 2014, the EU-funded PREPARE project (Platform foR European Preparedness Against (Re-) emerging Epidemics) was initiated to rapidly respond to severe infectious diseases outbreaks. It initiated two adaptive platform trials, the REMAP-CAP [3] and ALIC$^4$E trial [4], determining effectiveness of multiple treatment strategies for a single disease. REMAP-CAP was designed for a pandemic of severe community-acquired pneumonia and expanded globally during the COVID-19 pandemic. It has already delivered 10 important conclusions for better treatment of COVID-19 [3].

Antimicrobial resistance (AMR), on the other hand, is a more silent pandemic, which further augments the burden of infectious diseases. Development of new anti-bacterial agents has slowed, due to a combination of low return on investment and challenging clinical trial conditions [5]. Since the 2000s, national and global efforts have been initiated to promote antibiotic development, including the set-up of public-private partnerships, like COMBACTE (Combatting Bacterial Resistance in Europe), funded by EU's Innovative Medicines Initiative. From 2013, COMBACTE has focused on rapid and efficient development of new AMR prevention and treatment strategies, through site selection streamlining, real world data collection for patient enrichment and pathogen prioritization [6], and improvement of clinical trial efficiency through innovative methodological and statistical approaches [7].

A clinical research network, applying different innovative concepts, could enhance research efficiency to address both the growing threat of (re-)emerging infectious diseases and the increasing burden of AMR. Therefore, academic partners from PREPARE, COMBACTE, and other EU-funded collaborative networks (see acknowledgements), combined their strengths to establish the European Clinical Research Alliance on Infectious Diseases (Ecraid) Foundation in January 2022. A clinical research network that aims to efficiently generate rigorous evidence to improve the diagnosis, prevention and treatment of infections in Europe through innovative solutions. Here we will discuss the advantages and challenges of the implementation of Perpetual Observational Studies (POS).

*Methods to improve implementation efficiency of observational studies and randomized trials*

Various methods have been applied to improve the efficiency of clinical research. Clinical research networks have been set-up to improve research infrastructure; they provide essential resources to clinical researchers, including specialist training, information systems, administrative services, and communications expertise. Application of a master protocol takes efficiency one step further, as within a research infrastructure, a common protocol is applied to simultaneously coordinate multiple randomized controlled trials (RCTs) [8]. This structure can answer questions on one or more interventions in multiple diseases (basket trial), or multiple interventions in a single disease (umbrella or platform trial) [8]. In the latter case, umbrella trials stratify patients into subgroups based on (molecular) markers, but have fixed treatment arms, while platform trials have dynamic treatment arm assignment for a single indication [8]. RCTs can also be conducted within longitudinal epidemiological cohorts or within existing registries. In Trials Within Cohorts (TWICS), or cohort-multiple RCT (cmRCT), cohort participants consent to contribute control data to future unspecified trials at recruitment. Trial-specific consent is only requested from those participants randomly allocated to an intervention arm [9]. A POS combines important aspects of these different strategies; it is based in a research infrastructure, applies a core protocol, and entails a prospective cohort for epidemiological purposes. However, the POS research infrastructure has a broader focus (infectious diseases), its core protocol is more flexible, and, in contrast to TWICS, the POS can host multiple observational studies and RCTs in parallel. (table 1)

*What is a POS?*

A POS is a prospective, observational clinical study enrolling patients on a perpetual basis, collecting a set of demographics, clinical characteristics and outcomes, mostly available through routine care, as described in a core protocol. The POS in Ecraid are multi-centre studies designed to address key clinical research gaps, including variations in clinical practices, incidence of infectious disease syndromes and associated risk factors. Each core protocol establishes the general observational cohort and can be extended with appendices for more specific observational or interventional studies. Naturally, such 'add-on' studies would require resubmission for ethical and/or regulatory approval. Thus, each POS creates a clinical research backbone, ready to concurrently, or sequentially embed studies (observational, experimental, investigator-initiated or commercial) and efficiently advance the evidence-base for infectious diseases management. Within Ecraid, we have initiated five POS, recruiting patients in study sites across Europe, which target five strategically chosen infectious syndromes in hospital and community care settings (Table 2).

*What are the advantages of a POS within a clinical research network?*

Using the POS within a clinical research network, the quality, planning, and efficiency of experimental or observational multi-centre studies can be improved compared to conventional initiatives. Training and regular feedback improves data quality, creating a network of sites with experienced Principal Investigators and Good Clinical Practice-certified personnel.

For study planning, the POS collects empirical data, such as recruitment rates and cumulative incidence of endpoints of interest, which can: 1) help determine the feasibility of new studies within the domain and fine-tune their design, 2) improve site-selection for studies, especially for studies focused on specific pathogens or resistance traits; 3) inform sample-size calculations and timelines for embedded studies; 4) assess the feasibility and/or added value of implementing adaptive designs through empirically informed simulations.

Efficiency of embedded, multi-centre studies can be improved through shortened timelines for site contracting and study initiation via established professional relationships. Ethical and regulatory approvals can be expedited, as already approved documents can serve as templates for embedded studies.

The POS can also have other benefits. Clinical outcomes collected within the POS can be used for real-world data collection on newly registered drugs and vaccines. POS data from the full network can be used to assess external generalizability of RCT results, often based on a selective patient sample. A biobank of biological samples collected within the POS can provide a unique opportunity to study rare pathogens or resistance patterns. It takes research preparedness a step further, as the POS preemptively collects data that can promptly address arising research questions due to outbreaks, emerging infectious diseases, or pathogen variants. The clinical and epidemiological data can further serve to provide high-quality burden assessments for infectious syndromes, improving reliability and representativeness of existing estimates. Since the POS are integrated within a clinical research infrastructure, different additional services could be offered. Within Ecraid, expertise will be available for site selection (CLIN-Net and the Primary Care Network selection team), microbiology (LAB-Net), epidemiology (EPI-Net) and statistical methods (STAT-Net), depending on requirements of external study teams.

*Data harmonization and data sharing*

Data harmonization and data sharing are paramount for an effective clinical research network. This can be facilitated by standardized data collection in line with Clinical Data Acquisition Standards Harmonization (CDASH) principles [10], as well as through harmonization of diagnostic, treatment and follow-up procedures across clinical centres participating in the network. Finally, automated data collection through electronic health records could further enhance standardization, data quality and research efficiency, supporting long-term sustainability of the network. All these elements will be explored in the Ecraid-POS.

*Informed consent*

Since the POS are observational in nature, pose no risk to participants and serve a public benefit, a waiver of consent or opt-out procedure would be acceptable to many ethical committees. However, within ECRAID-Base it was decided to apply informed consent for all patients to optimize the POS benefits. It will ensure that POS data, like recruitment rates or frequency of clinical outcomes, will be more representative for future embedded RCTs, with an obligation of informed consent. It will also result in continuous training of patient recruitment and informed consent procedures, which will maximize enrolment of eligible patients in future embedded RCTs. Moreover, re-use of data and/or clinical samples is integral to the POS concept, for which broad consent for future use through an informed consent procedure is the preferred approach. Finally, the application of informed consent could improve reliability of post-discharge follow-up. However, informed consent also entails risk of bias related to non-participation and loss-to-follow-up due to consent withdrawal. Data from electronic medical records and the POS itself could be used to better understand and quantify this possible bias, and inform mitigation strategies.

*Challenges*

As the POS approach is new, careful explanation of the benefits, ethical issues, and quality safeguards will be necessary. While efficiency gains are expected, the speed of regulatory and ethical procedures will depend on the preparedness of ethical committees and regulatory authorities to facilitate these processes. Then, not all potential study sites may have the capacity to take on new, long-term studies, but the pool of possible study sites across Europe should be large enough to set-up efficacious POS networks. Moreover, POS focused on different infectious diseases can be run in parallel at the same site. Study site selection should be a transparent process, and participation in a POS should not limit a study site from participating in other research activities. Since a large pool of centres will be reporting data, data standardization will require detailed protocols, intensive training of sites and active data monitoring. Selection bias associated with informed consent will have to be monitored as well. Finally, demonstration of the benefits of POS will be required to keep researchers, sites and funders engaged, and to ensure long-term research sustainability.

*Conclusions*

The risk of (re-)emerging infectious diseases and the increasing prevalence of AMR requires a proactive stance to support efficient implementation of high-quality observational and interventional studies to better manage these health threats. The implementation of disease-specific POS within a clinical research network can improve the planning, quality and efficiency of multi-centre studies, and has the potential to shape the future of clinical research, although real-life benefits will need to be established. In Ecraid, five syndrome-specific POS are initiated, focused on VAP, cUTI, ARI in primary or secondary care, and unexplained severe infectious syndromes, which will provide important insights into the challenges and advantages of POS.

**Funding**

The Perpetual Observational Studies are part of ECRAID-Base. The ECRAID-Base project has received funding from the European Union's Horizon 2020 Research and Innovation programme, under the Grant Agreement number 965313.

**Acknowledgements**

We acknowledge the contribution of the different EU-funded collaborative networks, focusing on infectious diseases research, to the establishment of Ecraid, including COMBACTE, COMPARE, COMPARE-VEO, ECRIN, ISARIC, ORCHESTRA, PENTA, PREPARE, RECOVER, VACCELERATE, VALUE-Dx, VERDI, ReCoDID, ZIKAction, ZIKAlliance, and Zikaplan. We thank Frank Deege for his support and critical evaluation of the manuscript, and Ankur Krishnan and Lauren Maxwell for their input on ethical procedures.

**Conflicts of interest**

All authors are part of the ECRAID-Base consortium, the ECRAID-Base project has received funding from the European Union's Horizon 2020 Research and Innovation programme, under the Grant Agreement number 965313 to support the development of Perpetual Observational Studies.

**Author contributions**

MdK conceptualized and supervised writing of the commentary, NHK, HvW, TJ, JD and MdK drafted the manuscript and created the tables. All authors contributed to table 2, reviewed and edited the commentary and accepted the final version.

**Tables and figures**

Table 1. Advantages and challenges of different approaches that could improve efficiency of randomized clinical trial (RCT) implementation

| Approach | Description | Advantages | Challenges |
|---|---|---|---|
| Master protocol: Basket trial | RCT with one fixed treatment arm focused on multiple diseases or disease subtypes with the same molecular profile (mostly in oncology) | - Enables RCTs for rare diseases<br>- More efficient; can investigate multiple rare diseases within one RCT based on one master protocol | - Stratification and control per disease type is still required<br>- Challenging to select an endpoint, which is informative for all strata<br>- Limited to one treatment |
| Master protocol: Umbrella trial | RCT with multiple treatment arms for one disease, with subgroups based on (molecular) markers | - More efficient; can study multiple treatments within one RCT based on one master protocol, and utilizes one shared control arm | - Very challenging for pivotal trials; not easily accepted by regulatory authorities, collaboration of different pharmaceutical companies is required |
| Master protocol: Platform trial | RCT with multiple, dynamic treatment arms (in multiple domains) that can be dropped based on accumulating evidence, or added based on novel developments | - As above, more efficient<br>- Can minimize patients randomized to inferior treatment<br>- Very flexible | - As above challenging for pivotal trials<br>- Complex planning and implementation<br>- Possible influence of temporal trends that needs to be acknowledged |
| Registry based RCT | Patients part of a registry are randomized, after which documentation | - Efficient; limited interventions and follow-up outside of standard care | - Only possible for syndromes in settings with existing registries |

| | of treatment and/or outcomes occurs in the preexisting registry with outcome measurements at routine care time points | - High data quality through trained physicians<br>- Harmonized data through application of standardized definitions<br>- Less restrictive inclusion criteria; higher external validity<br>- Detailed information about non-participants<br>- Data available to inform design and sample size | - Little flexibility with regards to data collected, time of visits, or endpoint selection |
|---|---|---|---|
| Trial within cohort, cohort multiple RCT | Implementation of RCTs in existing disease-specific cohorts, with informed consent only for patients randomized to the experimental treatment arm | - Similar to registry based RCT<br>- Delayed informed consent for the treatment arm only, can reduce attrition and disappointment bias | - Disease-specific cohort needs to be established<br>- Standard treatment needs to be an acceptable comparator<br>- Non-response in patients refusing participation in treatment arm<br>- Important risk of cross-over from intervention to control arm |
| Perpetual observational study within a clinical research infrastructure | Network of study sites running a perpetual observational study based on a minimal protocol, ready for embedding | - Data available to inform center selection, study design and sample size<br>- Very flexible, additional data | - Organization of multiple parallel studies in the same network<br>- Motivation of participating sites |

|   | observational studies or RCTs | requirements can be implemented<br>- High data quality through trained personnel<br>- Harmonizated data through application of standardized definitions across studies<br>- Information available about non-participants<br>- Procedures and templates in place for site contracting and ethical approval<br>- Services available (site selection, microbiological, epidemiological and statistical expertise) | - Sustainability; Externally funded research activities need proper margins to sustain the POS activities<br>- Theoretical efficiency gains still need to be assessed in practice |
|---|---|---|---|

Table 2. Description of the five Perpetual Observational Studies initiated within the European Clinical Research Alliance on Infectious Diseases

| Name | Infectious syndrome | Healthcare setting | Number of anticipated sites | Number of anticipated patients per year | Description |
|---|---|---|---|---|---|
| **POS-ICU-VAP** | Ventilator-associated pneumonia | Intensive care unit | 40 | 4,000 | A platform for observational and randomized studies in the domains of VAP prevention, diagnosis, and |

| | | | | | treatment. Its primary endpoint is focused on study quality and efficiency, while its secondary endpoints focus on VAP epidemiology. Data collected in the POS will provide the information necessary for the design of any study in this field with regards to incidence and microbiological etiology of VAP and VAP outcomes including clinical or microbiological cure, duration of mechanical ventilation and mortality. |
|---|---|---|---|---|---|
| **POS—cUTI** | Complicated urinary tract infection | Hospital | 40 | 3,000 | A platform for rapidly implementing randomized studies focusing on cUTI. The POS will allow a continuous evaluation of best practices related to patient enrolment and data collection to improve study execution. It will also facilitate the harmonization of local practices in diagnosing and treating cUTI across study sites. The collection of clinical information will further support design of innovative clinical trials, including population enrichment. |
| **POS-ER-ARI** | Community-acquired acute respiratory tract infection | Secondary Care | 40 | 4,000 | This POS will be implemented in clinical networks that were active in PREPARE observational studies. It |

| | | | | | will provide an infrastructure capable of rapidly implementing clinical trials, related to the treatment of ARI. The POS will also answer key clinical questions on ARI, *e.g* effectiveness of different established diagnostic and therapeutic practices on clinical outcomes, and description of the burden of ARI presenting acutely to hospitals. |
|---|---|---|---|---|---|
| **POS-PC-ARI** | Community-acquired acute respiratory tract infection | Primary care | 50-100 | 2,000 registered, of which 400 included | This POS will be implemented in the Ecraid primary care research and PENTA networks. The PENTA sites will be recruiting children presenting to hospital-associated and out-of-hours care settings. Presentation and management details (diagnostics, prescribed medication) will be anonymously registered, with a subset of patients selected for capturing microbiological and outcome data. As a precursor to the POS, a point-prevalence audit survey was rolled out before, and three times during, the COVID-19 pandemic, showing this POS may become a valuable tool to capture shifts in |

|  |  |  |  |  | patient management due to changing circumstances. |
|---|---|---|---|---|---|
| **POS-ER-Disease X** | Infectious disease syndromes among immunocompromised patients admitted to hospital | Hospital | 5-8 | 400 | This POS aims to study febrile illness among immunocompromised adult patients admitted to the hospital with unknown or unusual viral aetiology. The network facilitates alignment of local practices of laboratory diagnosis among participating sites, and provides an infrastructure to evaluate novel diagnostic tools (i.e. metagenomics). |

POS, Perpetual Observational Study; VAP, ventilator-associated pneumonia; cUTI, complicated urinary tract infection; ARI, Acute Respiratory Infection: ER, Emergency Room; PREPARE, Platform foR European Preparedness Against (Re-) emerging Epidemics; RECOVER, Rapid European SARS-CoV-2 Emergency Research response; PC, Primary Care; PENTA, Paediatric European Network for Treatment of AIDS.